\documentclass[12pt,fleqn]{article}
\usepackage{epsf}
\usepackage{amssymb}
\begin{document}
%%%%%%%%%%%%%%%%%%%%%%%%%%%%%%%%%%%%%%%%%%%%%%

\title{Implications of Constraints on Mass Parameters in the Higgs Sector of
the Nonlinear Supersymmetric SU(5) Model}
\author{Dong Won Lee$^{(a,b)}$, Bjong Ro Kim$^{(b,c)}$, Sun Kun Oh$^{(a,b)}$
%,G. Kreyerhoff$^{(c)}$,
\\
\\
{\it $^{(a)}$ Department of Physics, Konkuk University, Seoul 143-701,
Korea} \\
{\it $^{(b)}$ CHEP, Kyungpook National University, Daegu 702-701, Korea} \\
{\it $^{(c)}$ III. Physikalisches Institut A, RWTH Aachen} \\
              {\it D-52056 Aachen, Germany} \\
\\
\\
}
\date{}
%\date{Received: 21.Feb.2001}
\maketitle
\thispagestyle{empty}

\newpage

\begin{abstract}
The Higgs sector of the minimal nonlinear supersymmetric SU(5) model contains three
mass parameters.
Although these mass parameters are essentially free at the electroweak scale,
they might have particular values if they evolve from a particular constraints
at the GUT scale through the RG equations.
By assuming a number of simple constraints on these mass parameters at the GUT scale,
we obtain their values at the electroweak scale through the RG equations in order
to investigate the phenomenological implications.
Some of them are found to be consistent with the present experimental data.
\end{abstract}

\vfil

%****************************************************************
\section{Introduction}
%****************************************************************

Although most of the popular supersymmetric models are linear ones,
it is still an open question whether supersymmetry is realized in nature
in linear or nonlinear way \cite{supersymmetry}.
One of us have considered a nonlinear realization of supersymmetry with SU(2) $\times$ U(1)
symmetry some years ago \cite{nonlinearSM}.
This model requires at least two Higgs doublets and a singlet for its Higgs sector.
Thus, at least with respect to the Higgs sector, this nonlinear model may be regarded as
an alternative to the linear Next-to Minimal Supersymmetric Standard Model (NMSSM).
Analysis of this nonlinear model show that it is consistent with phenomenology \cite{NMSSM}.

An unfavorable aspect of the NMSSM is that its Higgs sector is larger than the simplest
linear supersymmetric model, the well-known Minimal Supersymmetric Standard Model (MSSM),
which has just two Higgs doublets.
The nonlinear alternative that has the same Higgs sector as the MSSM is the minimal nonlinear
supersymmetric SU(5) model \cite{nonlinearSU5}.
The Higgs potential of the low energy limit of this nonlinear model needs effectively only
two Higgs doublets.
This model has been investigated in some detail by us \cite{nonlinearSU5,franz1,oneloop,rg}.

However, this minimal nonlinear supersymmetric SU(5) model has a disadvantage compared
to the MSSM.
That is, it has one more parameter than the MSSM:
The Higgs sector of this nonlinear model at the electroweak scale is determined by two
Higgs doublets, and the Higgs potential in terms of these Higgs doublets contains
in general three mass parameters.
These mass parameters are essentially free at the electroweak scale.
They are completely independent.
On the other hand, the Higgs sector of the MSSM has just two independent parameters.

Therefore, it is worthwhile to look for arguments which allow us to remove this
disadvantage, that is, to reduce the number of independent parameters.
One of the possibility is given by the {\it freedom of fine tuning},
that is, to impose some constraints on the mass parameters at the GUT scale.
If they are constrained at the GUT scale, their values at the electroweak scale would
no longer be free but determined by the renormalization group (RG) equations
that govern their evolutions as functions of energy scale.

In this article, we investigate the phenomenological implications of imposing constraints
on the mass parameters in the Higgs potential of the minimal nonlinear supersymmetric SU(5)
model.
By considering a number of simple constraints, which are in fact analogous to the various
constrained versions of the MSSM, we examine the mass of the lightest scalar Higgs boson,
as well as other Higgs bosons, and their production cross sections in $e^+e^-$ collisions.
We find that some of the constraints yield unphysical results or phenomenologically unacceptable
results whereas others give results that are consistent with the present experimental data.

This article is organized as follows: In the next section, we describe the argument for
the possibility of imposing constraints on the mass parameters.
In Section 3, we review the results of unconstrained Higgs potential.
In Section 4, we consider a number of constraints in the increasing order of complexity.
Among them we investigate three particular cases which are phenomenologically interesting.
Concluding discussions are given in the last section.

%**************************************************************
\section{The Higgs Potential without Parameters}
%**************************************************************

A peculiar aspect of the minimal nonlinear sypersymmetric SU(5) model in its
spontaneous symmetry breaking from SU(5) to SU(3)$\times$U(1)
is the necessity of {\it manifold fine tuning} in the following sense:
In the conventional SU(5) model the spontaneous symmetry breaking of
SU(5) to SU(3)$\times$U(1) is induced by the following vacuum expectation
values of the diagonal elements of the adjoint Higgs multiplet $H^{24}$
\begin{equation}
\langle H^{24} \rangle = V_G \left(\begin{array}{ccccc}
  2 & 0 & 0  & 0 & 0 \\
  0 & 2 & 0  & 0 & 0 \\
  0 & 0 & 2  & 0 & 0 \\
  0 & 0 & 0 & -3+\epsilon & 0 \\
  0 & 0 & 0 & 0 & -3+\epsilon
 \end{array} \right),
\end{equation}
where only one fine tuning parameter, $\epsilon$, is introduced, which is of order
$10^{-28}$ GeV, and $V_G$ is of order $10^{16}$ GeV.

In case of the minimal nonlinear supersymmetric SU(5) model, one needs to introduce
three fine tuning parameters such that the vacuum expectation value of $H^{24}$ is given by
\begin{equation}
\langle H^{24}\rangle = V_G \left(\begin{array}{ccccc}
  2+\epsilon_1 & 0  & 0  & 0 & 0 \\
  0  & 2+\epsilon_1 & 0  & 0 & 0 \\
  0  & 0  & 2+\epsilon_1 & 0 & 0 \\
  0  & 0  & 0  & -3+\epsilon_2 & 0 \\
  0  & 0  & 0  & 0 & -3+\epsilon_3
 \end{array} \right),
\end{equation}
where all of the three fine tuning parameters $\epsilon_1$, $\epsilon_2$, and $\epsilon_3$
are of order $10^{-28}$.
As they satisfy $3 \epsilon_1+\epsilon_2+\epsilon_3 = 0$, only two of them are independent.
We need fine tune them.
In the sense that the minimal nonlinear supersymmetric SU(5) model needs one more free
fine tuning parameter than the conventional SU(5) model, it might be said that  the former is
less natural than the latter, as far as the fine tuning is considered to be unnatural.

However, a remarkable merit  of the minimal nonlinear supersymmetric SU(5) model is that
there is a theoretically consistent method to break SU(5) to SU(3)$\times$U(1) with
no need of fine tuning.
Unfortunately, the result of the low energy limit of the minimal nonlinear supersymmetric SU(5)
model without fine tuning seems to be incompatible with existing experimental data,
which will be discussed shortly.

The Higgs potential of the minimal nonlinear supersymmetric SU(5) model,
after the breaking of SU(5) all the way down to SU(3)$\times$U(1),
in the low energy limit at the electroweak scale is given at the tree level by
\cite{nonlinearSU5,franz1}
\begin{eqnarray} \label{tree-potential}
        V &=& {1 \over 8} (g_1^2 + g_2^2) (|H_1|^2 - |H_2|^2)^2
            + {1 \over 2} g_2^2 |H_1^+ H_2|^2 \cr
        & &\mbox{}  + \lambda^2 (|H_1|^2 |H_2|^2
            - {1 \over 5} |H_1^T \epsilon H_2|^2)  \\
        & &\mbox{} + m_1^2 |H_1|^2 + m_2^2 |H_2|^2
            + m_3^2 (H_1^T \epsilon H_2 + {\rm h.c.}), \nonumber
\end{eqnarray}
where three mass parameters $m_1$, $m_2$ and $m_3$ are introduced.

These mass parameters are expressed as $m_i = C_i (V_G-\xi_i)$ ($i=1,2,3$), where $\xi_i$
is of the same order of $10^{16}$ GeV as $V_G$, and the dimensionless parameter $C_i$ is of
order of unity.
Generally, both $V_G$ and $\xi_i$ have to be fine tuned such that the difference
$V_G - \xi_i$ should be of order of electroweak scale in order to make the model suitable
for the electroweak phenomenology.
It turns out in the minimal supersymmetric SU(5) model that one can obtain without fine
tuning the mass parameters a theoretically consistent model as a low energy limit by
breaking first SU(5) to SU(3)$\times$SU(2)$\times$U(1) and then breaking dynamically
SU(3)$\times$SU(2)$\times$U(1) to SU(3)$\times$U(1).

First, the breaking of SU(5) to SU(3)$\times$SU(2)$\times$U(1) can be, as shown in
Ref \cite{nonlinearSU5}, accomplished by the vacuum expectation values of the quintuplets
$H^5$ and $\bar{H}^5$ as $\langle H^5 \rangle = 0$ and $\langle \bar{H}^5 \rangle = 0$,
respectively, and the vacuum expectation value of $H^{24}$ given independently of
$\epsilon_i$ as
\begin{equation}
\langle H^{24}\rangle = V_G \left(\begin{array}{ccccc}
  2  & 0  & 0  & 0 & 0 \\
  0  & 2  & 0  & 0 & 0 \\
  0  & 0  & 2  & 0 & 0 \\
  0  & 0  & 0  & -3  & 0 \\
  0  & 0  & 0  & 0 & -3
 \end{array} \right).
\end{equation}
The extremum conditions with respect to $\langle H^{24} \rangle$,
$\langle H^5 \rangle$ and $\langle \bar{H}^5 \rangle$ then imply that
the three mass parameters in the above tree-level Higgs potential are all zero:
$m_1 = m_2 = m_3 = 0$.

Now, for the Higgs potential with $m_1= m_2 = m_3 = 0$, if $\lambda = 0$,
SU(3)$\times$SU(2)$\times$U(1) is spontaneously broken to SU(3)$\times$U(1) at the tree level.
If, on the other hand, $\lambda \neq 0$, it is not possible for SU(3)$\times$SU(2)$\times$U(1)
to be spontaneously broken at the tree level but only possible at the one-loop level.
The parameters are evolved from the GUT scale to the electraoweak scale via the RG equations
given in Appendix A.
We carry out the  calculation in the frame of one-loop effective potential given in Appendix B.
The renormalization scale is taken to be between 100 GeV and 500 GeV.
It turns out that all loop contributions should be included: $b$ and $t$ quark, gauge bosons,
and scalar  Higgs bosons, where the masses of $b$ and $t$ quark, and the neutral gauge boson
are taken as $m_b = 4.3$ GeV, $m_t = 175$ GeV, and $m_Z = 91.187$ GeV, respectively.

Prior to the extremum conditions with respect to $\langle H_0\rangle$ and $\langle H_1\rangle$
are imposed, the Higgs potential has two independent parameters, namely,
$\lambda$ and $\tan\beta = v_2/v_1$.
After imposing the two extremum conditions, no free parameters are left in the Higgs potential.
Therefore, the Higgs potential contains no parameter; it may be called zero-parameter model.
For example the Higgs boson masses are uniquely fixed.
For $m_{S_1}$ we obtain 35 GeV.

In order to examine whether it is possible for these Higgs bosons to escape from experimental
detection, the production cross sections of $S_1$ in $e^+e^-$ collisions are calculated.
The relevant production channels are
\begin{equation}\label{productionchannel}
\begin{array}{l}
    {\rm (i)}~~ e^+ e^- \rightarrow Z \rightarrow ZS_i \rightarrow \bar{f} fS_i \cr
    {\rm (ii)}~~ e^+ e^- \rightarrow Z \rightarrow \bar{f}f \rightarrow \bar{f} fS_i \cr
    {\rm (iii)}~~ e^+ e^- \rightarrow Z \rightarrow PS_i \rightarrow \bar{f} fS_i \cr
    {\rm (iv)}~~ e^+ e^- \rightarrow \gamma \rightarrow \bar{f} f \rightarrow \bar{f} fS_i .
\end{array}
\end{equation}
For $\sqrt{s} = 92$ GeV, we obtain $\sigma_{S_1} = 7$ pb,
which is much larger than 1 pb, the discovery limit of LEP1.
Therefore, this zero-parameter model is phenomenologically incompatible with the LEP1 data,
although it is theoretically interesting in the sense that no fine tuning is required.

%*****************************************************************
\section{Unconstrained Higgs Potential}
%*****************************************************************

In this section, we summarize the results of unconstrained Higgs potential, where
the three mass parameters may have arbitrary values at the GUT scale,
in order to demonstrate the effects of the constraints \cite{franz1,oneloop,rg}.

The Higgs potential of the minimal nonlinear supersymmetric
SU(5) model contains three mass parameters, $m_i$ ($i=1,2,3$).
In general, the three mass parameters  $m_i = C_i (V_G-\xi_i)$ may take any
value between zero and say of order 1000 GeV.
If we do not use the {\it freedom of fine tuning}, the three mass parameters are not constrained.
In Ref. \cite{franz1}, the phenomenology of this unconstrained model has been treated at the
tree level.
In Ref. \cite{oneloop}, the analysis has been extended to the one-loop level in the frame
of effective potential method, where RG equations have not been used and only top and bottom
contributions have been taken into account.
In Ref. \cite{rg}, the RG equations have been derived and numerically solved in the
$\overline{\rm DR}$ scheme.
Evolving the parameters of the model from the GUT scale down to the electroweak scale,
the allowed regions of the parameters are determined, in particular the quartic coupling
constant $\lambda$.
The mass bounds, corrections to tree-level mass sum rules and productions of the Higgs
bosons at $e^+ e^-$
colliders are investigated for up to 2000 GeV of c.m. energy \cite{rg}.

We improve the results of these works by employing the RG equations given in Appendix A and
including not only top and bottom contributions but also gauge and Higgs self contributions
for the masses and cross sections.

At the GUT scale, we set the values of parameters to be
\begin{eqnarray}
    0 &\leq& \lambda_{\rm GUT} \leq 1.2  \cr
    -1 &\leq& m^2_{i_{\rm GUT}} \,\, (\rm TeV^2) \leq 1 \cr\
    1 &\leq& \tan\beta \leq 20,
\end{eqnarray}
where $i = 1, 2, 3$, and calculate their values at the electroweak scale using
the RG equations.
At the electroweak scale, we require the square masses of the Higgs bosons to be
positive and $\tan \beta$ to be in the range of $1 \leq \tan \beta \leq 20$.

We obtain the following numbers for $m_{S_1}, m_{S_2}$ and $m_P$,
which are respectively the masses of the Higgs scalars $S_1, S_2$ and pseudoscalar $P$:
\begin{eqnarray}
    31.6  \leq & m_{S_1}~(\rm GeV) & \leq 183.4 \cr
    114 \leq & m_{S_2}~(\rm GeV) & \leq 1311 \\
    24 \leq & m_P ~(\rm GeV) & \leq 1311. \nonumber
\end{eqnarray}

We calculate the production cross sections for the lightest scalar Higgs boson $S_1$ in
$e^+e^-$ collisions.
The relevant channels are the same as eq. (\ref{productionchannel}).
As no scalar Higgs boson has discovered at LEP, it might have escaped the detection or
its mass is bounded from below.
For  $\sqrt{s}= 205.9$  GeV, which is the center of mass energy reached finally at LEP2,
assuming the discovery limit of 40fb for LEP2, we find that $S_1$ should be heavier than
66 GeV in order to escape the detection at LEP2.

Future $e^+e^-$ colliders may discover the Higgs bosons of this unconstrained model.
Assuming that at least 10 signal events are necessary to detect the Higgs bosons, we set
the necessary minimum luminosity $L_{\rm min}$ for given center of mass energy of the
future $e^+e^-$ colliders:
For the $S_1$ production, we find that $L_{\rm min}$ is respectively
1.43 fb$^{-1}$, 5.4 fb$^{-1}$, and 21.3 fb$^{-1}$ for $\sqrt{s} = 500$, 1000, and 2000 GeV.
For the  $S_2$ production, we obtain that $L_{\rm min} = 23.8$  fb$^{-1}$ for
$\sqrt{s} = 2000$ GeV, and for the $P$ production, $L_{\rm min} = 77$ fb$^{-1}$
for $\sqrt{s} = 2000$ GeV.
An integrated luminosity of this order for the future linear collider is sufficiently
realistic, as the proposed linear collider project suggests that the baseline luminosity
for the $\sqrt{s} = 500$ GeV $e^+e^-$ linear collider is above 10$^{34}$ cm$^{-2}$s$^{-1}$
\cite{GLC}.

%****************************************************************
\section{Constrained Higgs Potential}
%****************************************************************

Now, let us use the {\it freedom of fine tuning} at the GUT scale.
The simplest form of fine tuning the mass parameters would be such that the number
of them is reduced.
In other words, we eliminate some of the mass parameters by fine tuning them.
%For example, we may eliminate all of them by tuning $m_1 = m_2 = m_3 = 0$, or
%two of them by setting either $m_1 = m_2 = 0$ but $m_3 \neq 0$, $m_2 = m_3 = 0$ but
%$m_1 \neq 0$, or $m_1 = m_3 = 0$ but $m_2 \neq 0$, and so on.
For example, we may eliminate all of them by tuning
$m^2_{1_{\rm GUT}} = m^2_{2_{\rm GUT}} = m^2_{3_{\rm GUT}} = 0$, or
two of them by setting either $m^2_{1_{\rm GUT}} = m^2_{2_{\rm GUT}} = 0$ but
$m^2_{3_{\rm GUT}} \neq 0$, $m^2_{2_{\rm GUT}} = m^2_{3_{\rm GUT}} = 0$ but
$m^2_{1_{\rm GUT}} \neq 0$, or $m^2_{1_{\rm GUT}} = m^2_{3_{\rm GUT}} = 0$
but $m^2_{2_{\rm GUT}} \neq 0$, and so on.
We find that among them, three cases of fine tunings yield phenomenologically
reasonable results:
\begin{equation}
\begin{array}{l}
%    ({\rm Case ~A}) ~ m^2_2 = 0, ~m^2_1 m^2_3 \neq 0  \cr
%    ({\rm Case ~B}) ~ m^2_1 = 0, ~m^2_2 m^2_3 \neq 0  \cr
%    ({\rm Case ~C}) ~ |m^2_1| = |m^2_2| = |m^2_3| \neq 0
%
    ({\rm Case ~A}) ~ m^2_{2_{\rm GUT}} = 0, ~m^2_{1_{\rm GUT}} \neq 0, ~m^2_{3_{\rm GUT}} \neq 0 \cr
    ({\rm Case ~B}) ~ m^2_{1_{\rm GUT}} = 0, ~m^2_{2_{\rm GUT}} \neq 0, ~m^2_{3_{\rm GUT}} \neq 0  \cr
    ({\rm Case ~C}) ~ |m^2_{1_{\rm GUT}}| = |m^2_{2_{\rm GUT}}| = |m^2_{3_{\rm GUT}}| \neq 0.
\end{array}
\end{equation}
We consider these three cases one by one.
Note that we take $0 \leq \lambda_{\rm GUT} \leq 1.2$ at the GUT scale for our
analysis and $0 < |m^2_{i_{\rm GUT}}|~\mbox{(TeV$^2$)} \leq 1$ for the mass parameters.
The other values we take in our calculations at the electroweak scale are
$m_b = 4.3$ GeV, $m_t = 175$ GeV, and $m_Z = 91.187$ GeV, and $1 \leq \tan \beta \leq 20$.

%****************************************************************
\subsection{Two-Parameter Higgs Potential}
%****************************************************************

{\bf Case A}

In Case A, there are two independent mass parameters.
We fine tune at the GUT scale one of the mass parameters to be zero, and
let the other two mass parameters vary independently.
From the GUT scale where we set $m^2_{2_{\rm GUT}} = 0$,
$0 < |m^2_{1_{\rm GUT}}|~\mbox{(TeV$^2$)} \leq 1$,
$0 < |m^2_{3_{\rm GUT}}|~\mbox{(TeV$^2$)} \leq 1$ and
$0 \leq \lambda_{\rm GUT} \leq 1.2$, the RG equations lead us at the electroweak
scale to
\begin{eqnarray}
    (107)^2  \leq & m_1^2~({\rm GeV}^2) & \leq  ~~~(1176)^2    \cr
    -(133)^2  \leq  & m_2^2~({\rm GeV}^2) & \leq  -(52)^2    \cr
    -(272)^2  \leq &  m_3^2~({\rm GeV}^2)  & \leq  -(45.4)^2    \\
    0.005  ~\leq  & \lambda  & \leq  ~~~0.388. \nonumber
\end{eqnarray}
With these allowed parameters, we calculate the Higgs boson masses at the electroweak
scale.
We plot $m_{S_1}$ in Fig. \ref{a-1}, where one can see that points are scattered between
104.6 GeV and 183.4 GeV for $m_{S_1}$. In this way, we set the ranges for the Higgs
boson masses.
The results are:
\begin{eqnarray}
    104.6 \leq & m_{S_1}~{\rm (GeV)} & \leq 183.4 \cr
    129.4 \leq & m_{S_2}~{\rm (GeV)} & \leq 1178 \\
    156 \leq & m_P ~{\rm (GeV)} & \leq 1178. \nonumber
\end{eqnarray}
Note that all the Higgs bosons are heavier than the $Z$ boson mass.
The allowed range for $m_{S_1}$ is rather tight compared to the allowed ranges for $m_{S_2}$
or $m_P$.

Now, the cross sections for the productions of these Higgs bosons are calculated in order
to check the possibility of detecting these Higgs bosons in $e^+e^-$ collisions.
For $\sqrt{s} = 205.9$ GeV, the center of mass energy of LEP2, the results of our calculations
show that the production cross sections for all these Higgs bosons are well below
the discovery limit of LEP2.
Thus, in Case A, the existing experimental data cannot put any constraints on the masses of
the three Higgs bosons in the minimal nonlinear supersymmetric SU(5) model.

The cross sections for the productions of these Higgs bosons at the future $e^+e^-$ linear
colliders are also calculated.
For $S_1$, we plot in Fig.\ref{a-500-s1} $\sigma_{S_1}$ for its production in $e^+e^-$
collisions at $\sqrt{s} = 500$ GeV.
One can see that $\sigma_{S_1}$ lies between about 7 and 9.8 fb.
We also calculate for other center of mass energies.
Thus, the results for $S_1$ production in  $e^+e^-$ collisions for
$\sqrt{s} = 500$ (1000, 2000) GeV are
\begin{equation}
    7 \,\,(1.85, \, 0.47) \leq \sigma_{S_1} ~{\rm (fb)} \leq 9.8 \,\, (2.4, \,\ 0.5).
\end{equation}

The lower bounds for $\sigma_{S_2}$ and $\sigma_P$ are nearly zero in  $e^+e^-$ collisions
at $\sqrt{s} = 500$ GeV.
This implies that they might not be discovered for some parameter regions of the minimal
nonlinear supersymmetric SU(5) model.
However, the upper bound for  $\sigma_{S_2}$ and $\sigma_P$ are comparatively larger than
that of $\sigma_{S_1}$: Our calculations yield that  $\sigma_{S_2} \le 285.1$ fb and
$\sigma_P \le 284.1$ fb at $\sqrt{s} = 500$ GeV.

In  $e^+e^-$ collisions at $\sqrt{s} = 1000$ GeV, both $S_2$ and $P$ might be produced heftily.
The production cross sections for both of them are obtained as
\begin{equation}
    0 \le  \sigma_{S_2, P} ~{\rm (fb)} \le 320,
\end{equation}
for $\sqrt{s} = 1000$ GeV.
Thus, in Case A, there are some parameter regions in the minimal nonlinear supersymmetric
SU(5) model where these Higgs bosons might be produced in large quantity at the future high
energy $e^+e^-$ linear colliders.

Extending our analysis for the future $e^+e^-$ linear colliders with $\sqrt{s} = 2000$ GeV,
we obtain that, as can be seen in Fig. \ref{a-lc2000s2p},
\begin{eqnarray}
    && \sigma_{S_2} \geq  1.9 ~{\rm fb} \cr
    && \sigma_P \geq 1.8 ~{\rm fb}.
\end{eqnarray}

{\bf Case B}

The Case B has also only two free mass parameters at the GUT scale.
We set  $m^2_{1_{\rm GUT}} = 0$, and allow other parameters to take values in the following ranges at
the GUT scale:
\begin{eqnarray}
    0 < & |m^2_{2_{\rm GUT}}|~\mbox{(TeV$^2$)} & \leq 1  \cr
    0 < & |m^2_{3_{\rm GUT}}|~\mbox{(TeV$^2$)} & \leq 1  \\
     0  \leq & \lambda_{\rm GUT} & \leq 1.2. \nonumber
\end{eqnarray}
Via RG equations, these parameters evolve from the GUT scale down to the electroweak scale
to have values as follows:
\begin{eqnarray}
    (40.6)^2  \leq & m_1^2~({\rm GeV}^2) & \leq  ~~(146.7)^2    \cr
    -(136.8)^2  \leq & m_2^2~({\rm GeV}^2) & \leq  -(61.7)^2   \cr
    -(94.4)^2  \leq & m_3^2~({\rm GeV}^2) & \leq ~~~~ 0   \\
    0.013 \leq & \lambda & \leq  ~~~0.388 . \nonumber
\end{eqnarray}
These values for the parameters yield relatively light Higgs bosons.
As are illustrated in Fig. \ref{b-ms1}, Fig. \ref{b-ms2} and Fig. \ref{b-mp}, respectively,
for $m_{S_1}$, $m_{S_2}$ and $m_P$, we obtain that
\begin{eqnarray}
31.6 \leq & m_{S_1}~{\rm (GeV)} & \leq 162 \cr
118 \leq & m_{S_2}~{\rm (GeV)} & \leq 191 \\
25.5 \leq & m_P ~{\rm (GeV)} & \leq 169 . \nonumber
\end{eqnarray}

With these mass ranges, $S_1$ and $P$ can be produced in $e^+e^-$ collisions at the center
of mass energy of LEP1, whereas $S_2$ production is not allowed kinematically.
However, the production of $P$ is suppressed due to the absence of its Higgs-strahlung process,
which is the dominant one for $S_1$ at the LEP1 energy.
So the no detection at LEP1 of $S_1$ may put a lower bound on $m_{S_1}$ as
\begin{equation}
    46 \le m_{S_1} ~{\rm (GeV)}.
\end{equation}

If the $e^+e^-$ center of mass energy is as large as the LEP2, all the three Higgs bosons can
be produced.
Here, too, the production of $P$ is strongly suppressed by the same reason as given at LEP1.
For LEP2 with $\sqrt{s} = 205.9$ GeV we plot $\sigma_{S_1}$ and $\sigma_{S_2}$
in Fig. \ref{b-lep206-ms1ms2}.
From this figure, assuming the discovery limit of 40 fb for LEP2, one might put a lower bound
on the mass of $S_1$ as $m_{S_1} \ge 67.5$ GeV.
On the other hand, one can see that $\sigma_{S_2}$ is smaller than 2 fb for the entire region
of the parameter space, which is well below the discovery limit of LEP2.
Thus, LEP2 cannot put any limit on $m_{S_2}$.

In $e^+e^-$ collisions with very high center of mass energy, the channel (iv) in equation
(\ref{productionchannel}) is comparably dominant with other channels in size and $\sigma_P$
becomes the same order of magnitude as $\sigma_{S_1}$ and $\sigma_{S_2}$.
We allow the parameters to vary within the ranges obtained by the RG equations, and calculate
the production cross sections.
We obtain the following lower bounds for them:
For $e^+e^-$ collisions with $\sqrt{s} = 500$ GeV,
\begin{eqnarray}
\sigma_{S_1} & \geq & 7 ~{\rm fb} \cr
\sigma_{S_2} & \geq & 6.4 ~{\rm fb} \\
\sigma_P & \geq & 2 ~{\rm fb} , \nonumber
\end{eqnarray}
for $e^+e^-$ collisions with $\sqrt{s} = 1000$ GeV,
\begin{eqnarray}
\sigma_{S_1} & \geq & 1.85 ~{\rm fb} \cr
\sigma_{S_2} & \geq & 1.5 ~{\rm fb}  \\
\sigma_P & \geq & 0.43 ~{\rm fb} , \nonumber
\end{eqnarray}
and for $e^+e^-$ collisions with $\sqrt{s} = 2000$ GeV,
\begin{eqnarray}
\sigma_{S_1} & \geq & 0.47 ~{\rm fb} \cr
\sigma_{S_2} & \geq & 0.42 ~{\rm fb}  \\
\sigma_P & \geq & 0.13 ~{\rm fb} . \nonumber
\end{eqnarray}

%****************************************************************
\subsection{One-Parameter Higgs Potential}
%****************************************************************

Let us consider the Case C.
The Higgs potential in the Case C contains only one mass parameter at the GUT scale,
namely, $0 < |m^2_{1_{\rm GUT}}| = |m^2_{2_{\rm GUT}}| =
|m^2_{3_{\rm GUT}}|\leq 1 \mbox{(TeV$^2$)}$ and we set $0 \leq \lambda_{\rm GUT} \leq 1.2$.
The RG equations yield their values at the electroweak scale as
\begin{eqnarray}
    (95)^2  ~\leq & m_1^2~({\rm GeV}^2) & \leq  ~~(296.7)^2    \cr
    -(110.6)^2 ~\leq  & m_2^2~({\rm GeV}^2) & \leq  -(113)^2   \cr
    -(214.3)^2 ~\leq  & m_3^2~({\rm GeV}^2) & \leq  -(56.9)^2   \\
    0.004 ~~\leq  & \lambda  & \leq  ~~~0.385 . \nonumber
\end{eqnarray}
And these values in turn yield the masses of the HIggs bosons as
\begin{eqnarray}
    85 \leq & m_{S_1}~{\rm (GeV)} & \leq 173 \cr
    141 \leq & m_{S_2}~{\rm (GeV)} & \leq 345 \\
    136 \leq & m_P ~{\rm (GeV)} & \leq 336 .
\nonumber
\end{eqnarray}

Now we calculate $\sigma_{S_1}$ at the LEP2 energy, $\sqrt{s} = 205.9$ GeV.
For given $m_{S_1}$, we search the entire region of the parameter space and select the largest $\sigma_{S_1}$.
In Fig. \ref{c-lep206-ms1}, we plot the result as a function of $m_{S_1}$.
Assuming the discovery limit of 40fb for LEP2, Fig. \ref{c-lep206-ms1} indicates that there are some parameter
regions for $m_{S_1} \le 107.3$ GeV where $S_1$ might be detected at LEP2.
Thus, the figure suggests that the lower bound on the mass of the lightest scalar Higgs boson in our model
is set as 107.3 GeV by LEP2.

In the future $e^+e^-$ linear colliders the cross section for the production of the
lightest scalar Higgs boson $S_1$ in this case is
\begin{equation}
    7.5 \,\,(1.9, \, 0.48) \leq \sigma_{S_1} ~{\rm (fb)} \leq 12.5 \,\, (4.3, \,\ 1.18) ,
\end{equation}
for $\sqrt{s} = 500$ (1000, 2000) GeV. For other Higgs bosons, we obtain that
\begin{eqnarray}
    && 0 \,\,(1.5, \, 0.42) \leq \sigma_{S_2} {\rm (fb)} \leq 80 \,\, (35, \, 8) \cr
    && 0 \,\,(1.0, \, 0.28) \leq \sigma_{P} {\rm (fb)} \leq 78 \,\, (35, \, 8) ,
\end{eqnarray}
for $\sqrt{s} = 500$ (1000, 2000) GeV.
The tendency is that the cross sections decrease as the Higgs bosons become heavier
and the cross sections increase as the Higgs bosons become lighter.
For $S_2$ and $P$, the minimum cross section for producing them
at a $\sqrt{s} = 500$ GeV $e^+e^-$ colliding machine is nearly zero.
However, the large upper bounds on the production cross sections suggest
that they might also be detected at the future $e^+e^-$ linear colliders depending
on their masses.

Note that these numbers are large enough for the future $e^+e^-$ linear colliders to
examine the Case C of the minimal nonlinear supersymmetric SU(5) model.
Thus, if the discovery limit for the $e^+e^-$ linear colliders at $\sqrt{s} = 500$ GeV
is 10 events, one would need an integrated luminosity of at least about 1.33 fb$^{-1}$.
In order to test the model by detecting, for example, $S_1$,  the minimum cross section of
whose production is about 7.5 fb.

%****************************************************************
\section{Discussions and Conclusions}
%****************************************************************

We have investigated if the minimal nonlinear supersymmetric SU(5) model is
phenomenologically viable, by fine-tuning the mass parameters of the Higgs potential.
We have set some of the mass parameters to be constrained at the GUT scale,
and then have evolved them down to the electroweak scale via RG equations.
We have found that three cases emerge as acceptable.

One of them is the case where $m_2$, the mass term of the Higgs
doublet $H_2$, which gives mass to the top-quark, is set to be zero at the GUT scale.
A characteristic feature of this case is that the mass of the lightest scalar Higgs boson
is predicted as 104.6 $\leq m_{S_1} ~(\rm {GeV}) \leq$ 183.4.
Note that the lower bound of $m_{S_1}$ is rather large while the allowed range of $m_{S_1}$
is comparatively narrow.

Another case is obtained by fine-tuning $m_1$, the mass term of the doublet
$H_1$, which gives mass to the bottom-quark, to be zero at the GUT scale.
A novel feature of this case is that all scalar Higgs bosons are predicted to be
lighter than 200 GeV.

The other case has only one mass parameter at the GUT scale.
It is obtained by fine-tuning $|m^2_{1_{\rm GUT}}|$, $|m^2_{2_{\rm GUT}}|$ and
$|m^2_{3_{\rm GUT}}|$ to be equal non-zero value at the GUT scale.
In this case, all scalar Higgs bosons are predicted to be between 85 GeV and 345 GeV.

We have also shown that these three cases are compatible with the data of LEP1 and LEP2.
We have calculated the lower bounds for the production cross sections of some Higgs bosons
at the future $e^+e^-$ colliders with $\sqrt{s} = 500$, 1000, and 2000 GeV.
The numbers are within the range of the discovery limit of the future machines,
thus allowing our model to be examined.

%******************************************************************
\newpage

\noindent{\Large \bf  Appendix A \\
\\
RG equations of the nonlinear Supersymmetric SU(5) model}
\\
\\
The RG equations of the parameters of our model are derived as follows:
\begin{eqnarray}
{d \lambda_1 \over d t } &=& {1 \over 16 \pi^2} \Big\{12 \lambda_1^2 +4 \lambda_3^2
+4 \lambda_3 \lambda_4 +2\lambda_4^2 +2\lambda_5^2 + 24\lambda_6^2 \cr
& &\mbox{}~~~~~~~~ - \lambda_1 (3 g_1^2 +9 g_2^2 -12 h_b^2)
+{3 \over 2} g_2^4 \cr
& &\mbox{}~~~~~~~~ +{3 \over 4} (g_1^2+ g_2^2)^2 -12 h_b^4 \Big\} \cr
{d\lambda_2\over d t }
&=& {1 \over 16 \pi^2} \Big\{12 \lambda_2^2 +4 \lambda_3^2 +4
\lambda_3 \lambda_4 +2\lambda_4^2 +2\lambda_5^2 + 24\lambda_6^2 \cr
& &\mbox{}~~~~~~~~ - \lambda_2 (3 g_1^2 +9 g_2^2 -12 h_t^2)
+{3 \over 2} g_2^4 \cr
& &\mbox{}~~~~~~~~ +{3 \over 4} (g_1^2+ g_2^2)^2 -12 h_t^4 \Big\} \cr
{d\lambda_3 \over d t }
&=& {1 \over 16 \pi^2} \Big\{4 \lambda_3^2 +2 \lambda_4^2 +
(\lambda_1 +\lambda_2) (6 \lambda_3 +2 \lambda_4) \cr
& &\mbox{}~~~~~~~~ + 2\lambda_5^2 + 4\lambda_6^2 +4\lambda_7^2 + 16\lambda_6\lambda_7 \cr
& &\mbox{}~~~~~~~~ - \lambda_3 (3 g_1^2 +9 g_2^2 -6 h_b^2-6 h_t^2) \cr
& &\mbox{}~~~~~~~~ +{9 \over 4} g_2^4 +{3 \over 4}g_1^4 -12 h_b^2 h_t^2 \Big\} \cr
{d \lambda_4 \over d t } &=& {1 \over 16 \pi^2}
\Big\{8\lambda_3 \lambda_4 +2 \lambda_4 (\lambda_1+\lambda_2) +4 \lambda_4^2
+8\lambda_5^2 \cr
& &\mbox{}~~~~~~~~ + 10\lambda_6^2 + 10\lambda_7^2 + 4\lambda_6\lambda_7 \cr
& &\mbox{}~~~~~~~~ - \lambda_4 (3 g_1^2 +9 g_2^2 -6 h_b^2 -6 h_t^2) \cr
& &\mbox{}~~~~~~~~ +12 h_b^2 h_t^2 +3 g_1^2 g_2^2 \Big\} \cr
{d \lambda_5 \over d t } &=& {1 \over 16 \pi^2}
\Big\{ 2\lambda_5 (\lambda_1+\lambda_2+4\lambda_3+6\lambda_4)
+10(\lambda_6^2+\lambda_7^2) \cr
& &\mbox{} ~~~~ +4\lambda_6\lambda_7
-\frac{1}{2}\lambda_5(18g_2^2+6g_1^2-12(h_t^2+h_b^2))
\Big\} \cr
{d \lambda_6 \over d t } &=& {1 \over 16 \pi^2}
\Big\{ 2\lambda_6 (6\lambda_1+3\lambda_3+4\lambda_4+5\lambda_5) \cr
& &\mbox{}~~~~~~~~ +2\lambda_7(3\lambda_3+2\lambda_4+\lambda_5) \cr
& &\mbox{}~~~~~~~~ -\frac{1}{4}\lambda_6(36g_2^2+12g_1^2-12(h_t^2+3h_b^2))
\Big\} \cr
{d \lambda_7 \over d t } &=& {1 \over 16 \pi^2}
\Big\{ 2\lambda_7 (6\lambda_2+3\lambda_3+4\lambda_4+5\lambda_5) \cr
& &\mbox{}~~~~~~~~ +2\lambda_6(3\lambda_3+2\lambda_4+\lambda_5) \cr
& &\mbox{}~~~~~~~~ -\frac{1}{4}\lambda_7(36g_2^2+12g_1^2-12(3h_t^2+h_b^2))
\Big\} \cr
{d \mu_1^2 \over d t} &=& {1 \over 32 \pi^2}
\Big\{12 \lambda_1 \mu_1^2 +(8 \lambda_3 +4 \lambda_4) \mu_2^2 \cr
& &\mbox{} ~~~~+24\lambda_6 \mu_3^2 -2(9g_2^2+3g_1^2-12h_b^2)\mu_1^2
\Big\} \cr
{d \mu_2^2 \over d t} &=& {1 \over 32 \pi^2} \Big\{12 \lambda_2 \mu_2^2
+(8 \lambda_3 +4 \lambda_4) \mu_1^2 \Big\} \cr
& &\mbox{} ~~~~+24\lambda_7 \mu_3^2 -2(9g_2^2+3g_1^2-12h_t^2)\mu_2^2
\Big\} \cr
{d \mu_3^2 \over d t} &=& {1 \over 32 \pi^2} \Big\{
(4 \lambda_3 +8\lambda_4 +12\lambda_5) \mu_3^2
+12\lambda_6 \mu_1^2 \cr
& &\mbox{} ~~+ 12\lambda_7 \mu_2^2 - (18g_2^2+6g_1^2 -12h_b^2 -12 h_t^2)\mu_3^2
\Big\} \cr
{d h_t \over d t} &=& - { h_t \over 16 \pi^2} \left(
8 g_3^2+{9 \over 4} g_2^2 + {17 \over 12}g_1^2 -{1 \over 2} h_b^2
- {9 \over2} h_t^2 \right)  \cr
{d h_b \over d t} &=& - { h_b \over 16 \pi^2} \left(
8 g_3^2+{9 \over 4} g_2^2 + {5 \over 12}g_1^2 -{1 \over 2} h_t^2
- {9 \over2} h_b^2 \right) ,
\end{eqnarray}
where the following redefinition of the parameters are used
\begin{eqnarray}
    \lambda_1 (M_{\rm GUT})
        &=& { g^2_1 (M_{\rm GUT}) +g^2_2 (M_{\rm GUT}) \over 4}  \cr
    \lambda_2 (M_{\rm GUT})
        &=& { g^2_1(M_{\rm GUT}) +g^2_2 (M_{\rm GUT}) \over 4}  \cr
    \lambda_3 (M_{\rm GUT})
        &=& { g^2_2 (M_{\rm GUT}) -g^2_1(M_{\rm GUT}) \over 4}
                        +\lambda^2 (M_{\rm GUT})  \cr
    \lambda_4 (M_{\rm GUT})
        &=& -{1 \over 2} g^2_2(M_{\rm GUT}) - {1 \over 5}
                        \lambda^2 (M_{\rm GUT})  \cr
    \lambda_5 (M_{\rm GUT})
        &=& \lambda_6 (M_{\rm GUT}) = \lambda_7 (M_{\rm GUT}) = 0 \cr
    \mu_i ^2 (M_{\rm GUT}) &=& m_i^2 (M_{\rm GUT}),
\nonumber
\end{eqnarray}
where $i = 1, 2, 3$.
Note that $\mu_1$ and $\mu_2$ are eventually eliminated from the potential
by the extremum conditions.

From the known values of the gauge couplings at $m_Z$ scale \cite{datagroup}
we obtain $g_1^2(m_Z) = 0.1283$, $g_2^2(m_Z) = 0.4273$ and $g_3^2(m_Z)
= 1.4912$ in the $\overline{\rm DR}$ renormalization scheme.
Through their RG evolution from $m_Z$ scale to $m_t$ scale with five quarks
and one Higgs doublet, the top-quark Yukawa coupling is obtained from
$m_t^{\rm pole} = {1\over\sqrt{2}}h_t(m_t) v_2(m_t) \\
(1+{5\over 3\pi} \alpha_s(m_t))$
in the $\overline{\rm DR}$ renormalization scheme at $m_t = 175$ GeV
\cite{mtpole}, where for the evolution of the gauge couplings we use
their two-loop $\beta$-functions \cite{twoloopbeta}.
In this way, the values of the gauge and the Yukawa couplings at $M_{\rm GUT}$ scale are
obtained using RG equations.
Then by applying these values and the remaining input parameters, $\lambda$
and $m_i^2$, as the boundary conditions for the RG equations at $M_{\rm GUT}$ scale,
the numerical values of the relevant parameters at the electroweak scale are
obtained through the RG evolution from $M_{\rm GUT}$ scale.
\\
\\

\noindent{\Large \bf Appendix B \\
\\
One-loop effective Higgs potential}
\\
\\
The effective potential $V_{\rm eff}$ at the one-loop level may conveniently
be decomposed as
\begin{equation}
        V_{\rm eff} =V_0 + V_b + V_t + V_g + V_s,
\end{equation}
where $V_0$ denotes the tree-level potential, the equation (\ref{tree-potential}),
$V_b(V_t)$ the $b$-quark ($t$-quark) contribution, $V_g$ the gauge boson
contribution, and $V_s$ the contribution of the Higgs bosons. As
for $V_s$ we first calculate the full field-dependent squared mass
matrix, then omit terms containing charged Higgs fields, which do
not contribute to the physical mass matrix of the Higgs bosons.
Then $V_s$ can be expressed as a sum of $V_{sc}$ and $V_{sn}$,
whereby $V_{sc}$ is the contribution of the field-dependent squared mass
matrix of the charged Higgs bosons, $V_{sn}$ that of the neutral Higgs bosons.
They are given by
%\begin{eqnarray}
%        V_0 &=& {1 \over 8} (g_1^2 + g_2^2) (|H_1|^2 - |H_2|^2)^2
%            + {1 \over 2} g_2^2 |H_1^{\dag} H_2|^2
%            + \lambda^2 (|H_1|^2 |H_2|^2
%            - {1 \over 5} |H_1^T \epsilon H_2|^2)  \cr
%        & &\mbox{} + m_1^2 |H_1|^2 + m_2^2 |H_2|^2
%            + m_3^2 (H_1^T \epsilon H_2 + {\rm h. c.})
%\end{eqnarray}
%where $g_1$ and $g_2$ are the coupling constants of U(1) and SU(2),
%respectively, $\lambda$ the dimensionless self coupling constant which comes
%from the coupling of Higgs quintuplets at the GUT scale \cite{franz1},
%$m_1^2$, $m_2^2$, $m_3^2$ the mass parameters, and $H_1^T = (H_1^0 , H_1^-)$
%and $H_2^T = (H_2^+ , H_2^0)$ Higgs doublets, and
\begin{eqnarray} \label{higgs-self-eff}
       V_b &=& - {3 {\cal M}_b^4 \over 16 \pi^2}
        \left (\log { {\cal M}_b^2  \over \mu^2} - {3 \over 2}
        \right )  \cr
        V_t &=& - {3 {\cal M}_t^4 \over 16 \pi^2}
        \left (\log {{\cal M}_t^2  \over \mu^2} - {3 \over 2}
        \right )  \cr
       V_g &=& {3 {\cal M}_W^4 \over 32 \pi^2}
        \left ( \log {{\cal M}_W^2 \over \mu^2} - {3 \over 2}  \right)
         + {3 {\cal M}_Z^4 \over 16 \pi^2}
        \left ( \log { {\cal M}_Z^2 \over \mu^2} - {3 \over 2} \right)  \cr
        \cr
     V_s &=&  V_{sc} + V_{sn} \cr
        \cr
%%%%%%%%%%%%%%%%%%%%%%%%%%%%%%%%%%%%%%%%%%%%%%%%%
       V_{sc} &=& {1 \over 32 \pi^2}
         Str \left [{\cal M}_C^2 {\cal M}_C^2 \left \{\log \left
         ({{\cal M}_C^2 \over \mu^2} \right)
        - {3 \over 2}\right \}    \right] \cr
%%%%%%%%%%%%%%%%%%%%%%%%%%%%%%%%%%%%%%%%%%%%%%%%%
        V_{sn} &=& {\sum_{{\cal H}=S,P} \over 64 \pi^2}
        Str \left [{\cal M}_{\cal H}^2 {\cal M}_{\cal H}^2 \left \{
        \log \left ({{\cal M}_{\cal H}^2 \over \mu^2} \right) - {3 \over 2}\right \}
    \right ],
\end{eqnarray}
where $\mu$ the renormalization scale and ${\cal M}$ denote the field-dependent
mass matrices for the particles \cite{rg}.
The Higgs doublets of the potential $V_0$, the equation (\ref{tree-potential}), can
be defined as follows
\begin{eqnarray}
        H_1 = \left(\begin{array}{c}
        {1 \over \sqrt {2}} (S_1 + i P_1) \\
                                H_1^-
        \end{array} \right) ,~
        H_2 = \left(\begin{array}{c}
                H_2^+ \\
                {1 \over \sqrt{2}} (S_2 + i P_2)
        \end{array} \right) .
\end{eqnarray}

%*****************************************************************
\vskip 0.3 in

\noindent
{\Large {\bf Acknowledgments}}

B. R. Kim was supported by the Brain Pool Program of KOSEF. He would like to
thank the hospitality of CHEP, Kyungpook National University.
%\vskip 0.3 in
\vfil\eject

\vspace{1cm}
%******************************************************************

\vfil\eject
%*************************************************************************

%***********************************************************************************
%Figure Captions
%***********************************************************************************
\newpage

{\bf Figure Captions}
\vskip 0.3 in

\begin{figure}[ht]
\begin{center}
%\vspace*{-0.5cm}
\mbox{\epsfxsize=8cm \epsfysize=8cm
\epsffile{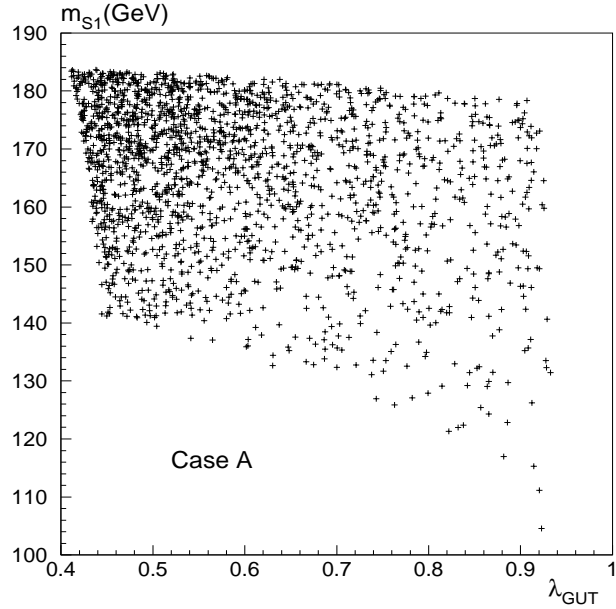}}
%\centerline{\includegraphics*[scale=0.5]{/@thesis/ch3fig/fig2a.eps}}
\caption{The plot of the RG-improved mass at the one-loop level of $S_1$
against $\lambda_{\rm GUT}$, for $0 \leq \lambda_{\rm GUT} \leq 1.2$,
$m^2_{2_{\rm GUT}} = 0$, $0 < |m^2_{1_{\rm GUT}}|~\mbox{(TeV$^2$)} \leq 1$ and
$0 < |m^2_{3_{\rm GUT}}|~\mbox{(TeV$^2$)} \leq 1$ at the GUT scale.}
\label{a-1}
%\vspace*{-1.cm}
\end{center}
\end{figure}

\noindent
Fig. \ref{a-1}: \ The plot of the RG-improved mass at the one-loop level of $S_1$
against $\lambda_{\rm GUT}$, for $0 \leq \lambda_{\rm GUT} \leq 1.2$,
$m^2_{2_{\rm GUT}} = 0$, $0 < |m^2_{1_{\rm GUT}}|~\mbox{(TeV$^2$)} \leq 1$ and
$0 < |m^2_{3_{\rm GUT}}|~\mbox{(TeV$^2$)} \leq 1$ at the GUT scale.
\vskip 0.2 in

\begin{figure}[hb]
\begin{center}
\vspace*{-0.5cm}
\mbox{\epsfxsize=8cm \epsfysize=8cm
\epsffile{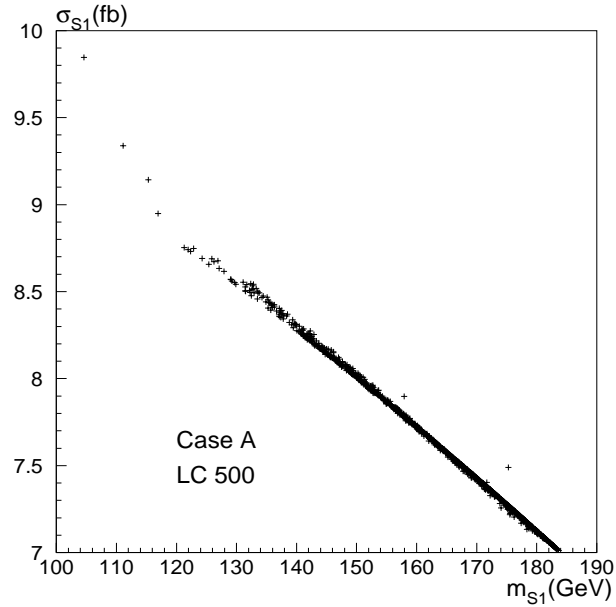}}
%\centerline{\includegraphics*[scale=0.5]{/@thesis/ch3fig/fig2a.eps}}
\caption{The plot against $m_{S_1}$ of $S_1$ production cross sections
at the one-loop level with the RG-improved effective potential
$V_{\rm eff}$ at future $e^+e^-$ collider for $\sqrt{s} = 500 ~{\rm GeV}$
in the Case A.}
\label{a-500-s1}
%\vspace*{-0.3cm}
\end{center}
\end{figure}

\noindent
Fig. \ref{a-500-s1}: \ The plot against $m_{S_1}$ of $S_1$ production cross sections
at the one-loop level with the RG-improved effective potential
$V_{\rm eff}$ at future $e^+e^-$ collider for $\sqrt{s} = 500 ~{\rm GeV}$
in the Case A.
\vskip 0.2 in

\begin{figure}[h]
\begin{center}
\vspace*{-0.cm}
\mbox{\epsfxsize=8cm \epsfysize=8cm
\epsffile{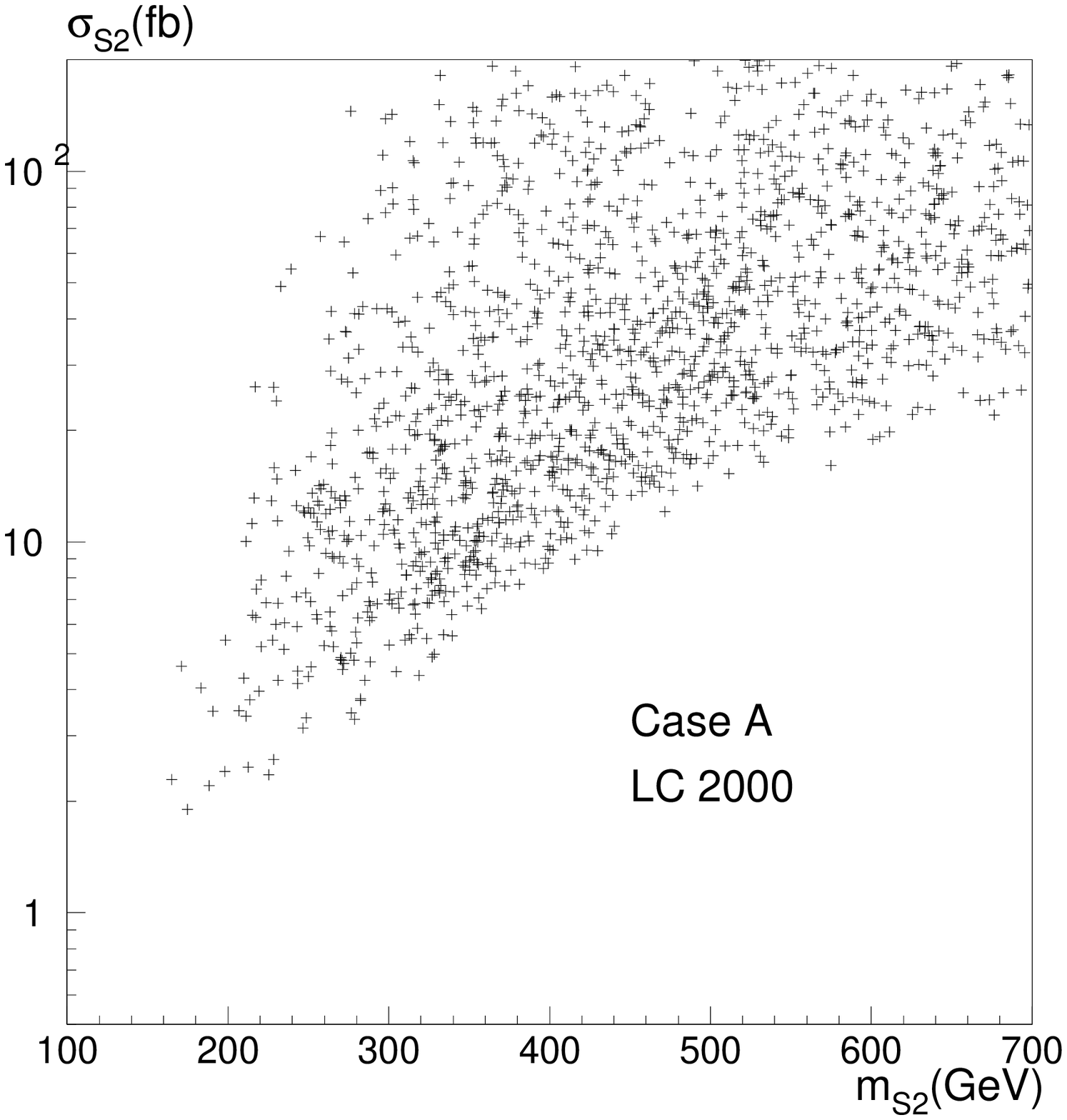}} \\
%\vspace*{.5cm}
\mbox{\epsfxsize=8cm \epsfysize=8cm
\epsffile{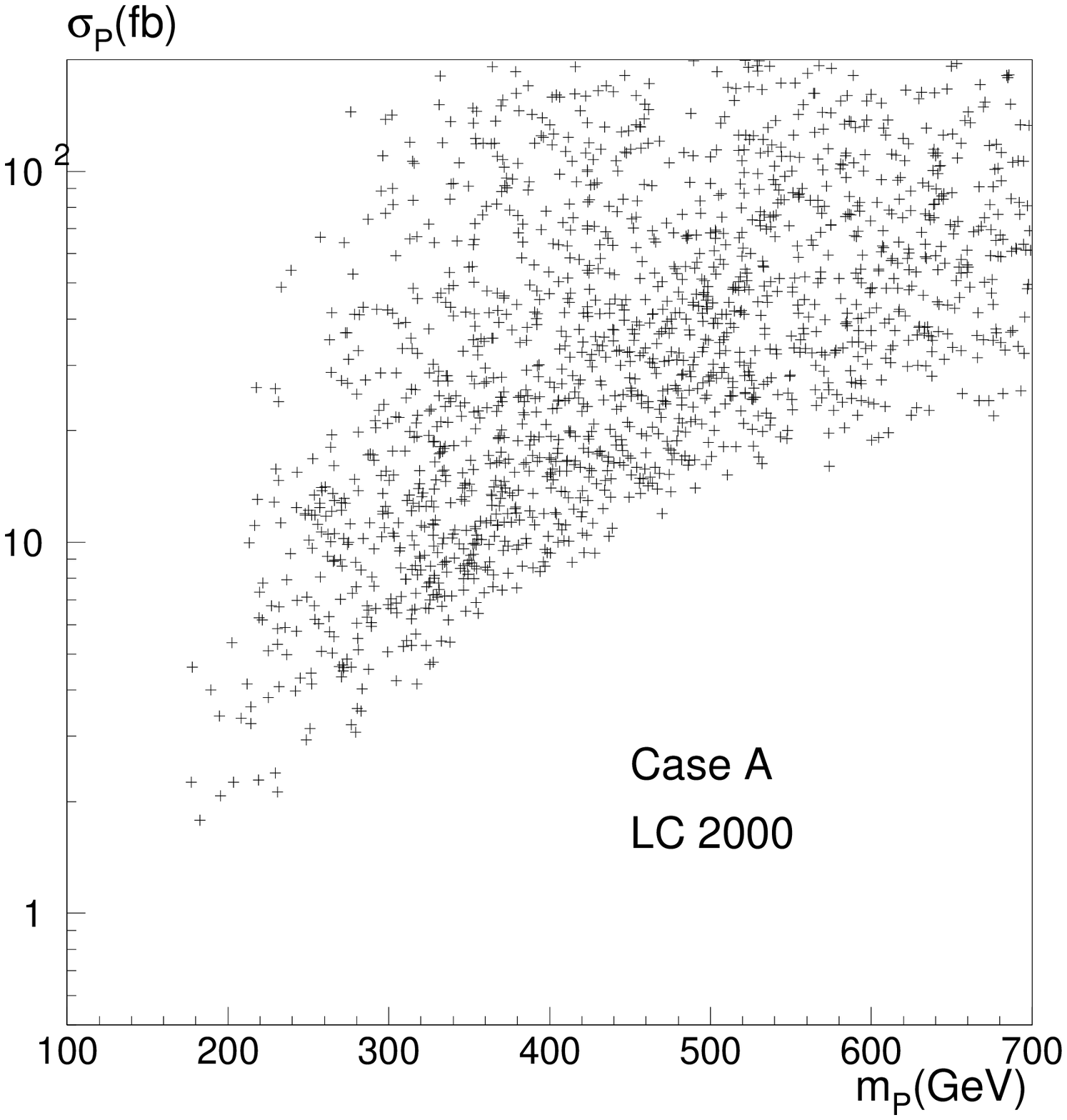}}
%\vspace*{0.4cm}
%\centerline{\includegraphics*[scale=0.5]{/@thesis/ch3fig/fig2a.eps}}
\caption{The plot against $m_{S_2}$ and $m_P$ of $\sigma_{S_2}$ and
$\sigma_P$, respectively, at the one-loop level with the RG-improved
effective potential $V_{\rm eff}$ at future $e^+e^-$ collider for
$\sqrt{s} = 2000 ~{\rm GeV}$ in the Case A.}
\label{a-lc2000s2p}
%\vspace*{-1.3cm}
\end{center}
\end{figure}

\noindent
Fig. \ref{a-lc2000s2p}: \ The plot against $m_{S_2}$ and $m_P$ of $\sigma_{S_2}$ and
$\sigma_P$, respectively, at the one-loop level with the RG-improved
effective potential $V_{\rm eff}$ at future $e^+e^-$ collider for
$\sqrt{s} = 2000 ~{\rm GeV}$ in the Case A.
\vskip 0.2 in

\begin{figure}[hp]
\begin{center}
%\vspace*{-0.0cm}
\mbox{\epsfxsize=8cm \epsfysize=8cm
\epsffile{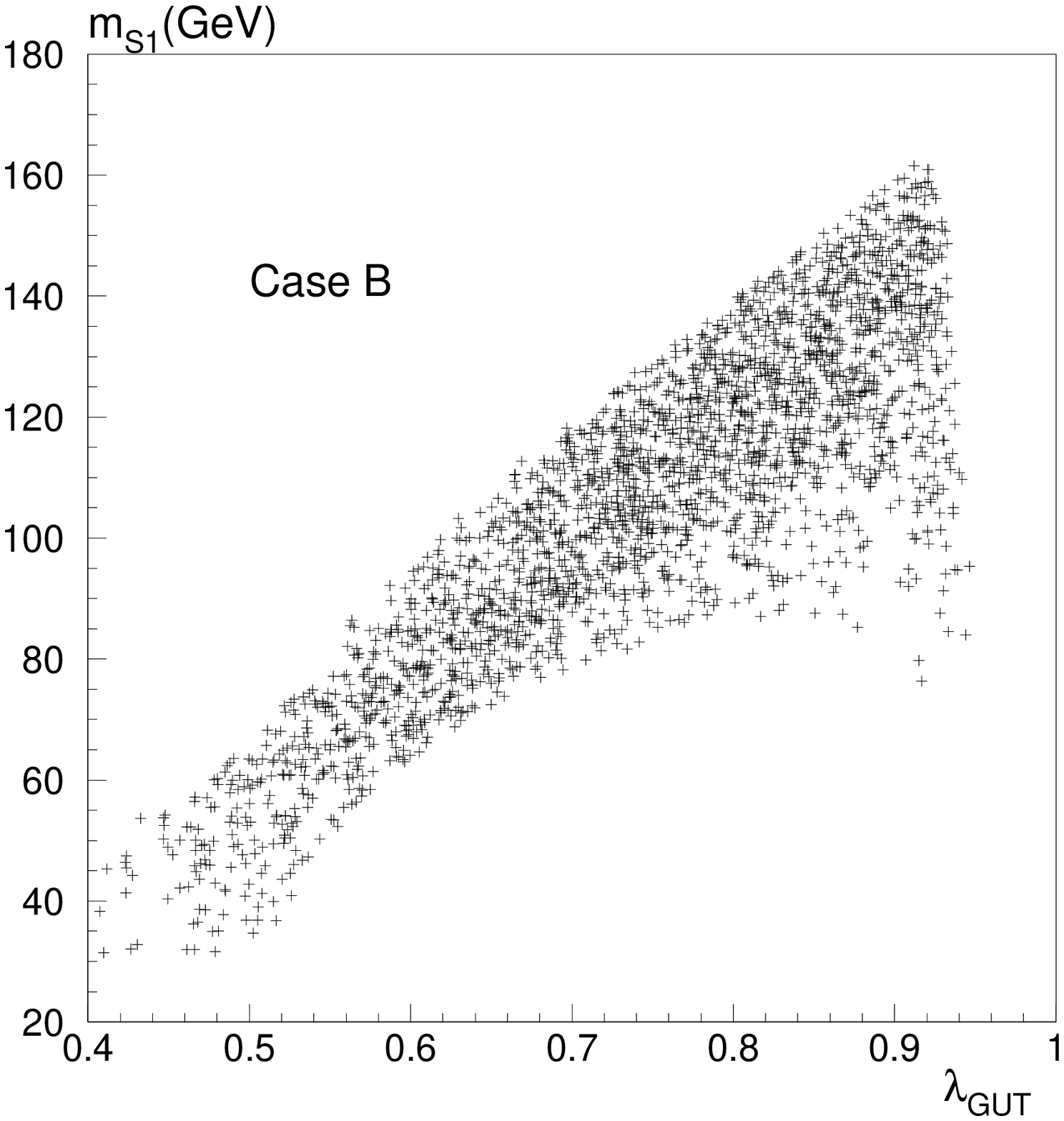}}
\caption{The plot of the RG-improved mass at the one-loop level of $S_1$
against $\lambda_{\rm GUT}$, for $0 \leq \lambda_{\rm GUT} \leq 1.2$,
$m^2_{1_{\rm GUT}} = 0$, $0 < |m^2_{2_{\rm GUT}}|~\mbox{(TeV$^2$)} \leq 1$ and
$0 < |m^2_{3_{\rm GUT}}|~\mbox{(TeV$^2$)} \leq 1$ at the GUT scale.}
\label{b-ms1}
\end{center}
\end{figure}

\noindent
Fig. \ref{b-ms1}: \ The plot of the RG-improved mass at the one-loop level of $S_1$
against $\lambda_{\rm GUT}$, for $0 \leq \lambda_{\rm GUT} \leq 1.2$,
$m^2_{1_{\rm GUT}} = 0$, $0 < |m^2_{2_{\rm GUT}}|~\mbox{(TeV$^2$)} \leq 1$ and
$0 < |m^2_{3_{\rm GUT}}|~\mbox{(TeV$^2$)} \leq 1$ at the GUT scale.
\vskip 0.2 in

\begin{figure}[hb]
\begin{center}
\mbox{\epsfxsize=8cm \epsfysize=8cm
\epsffile{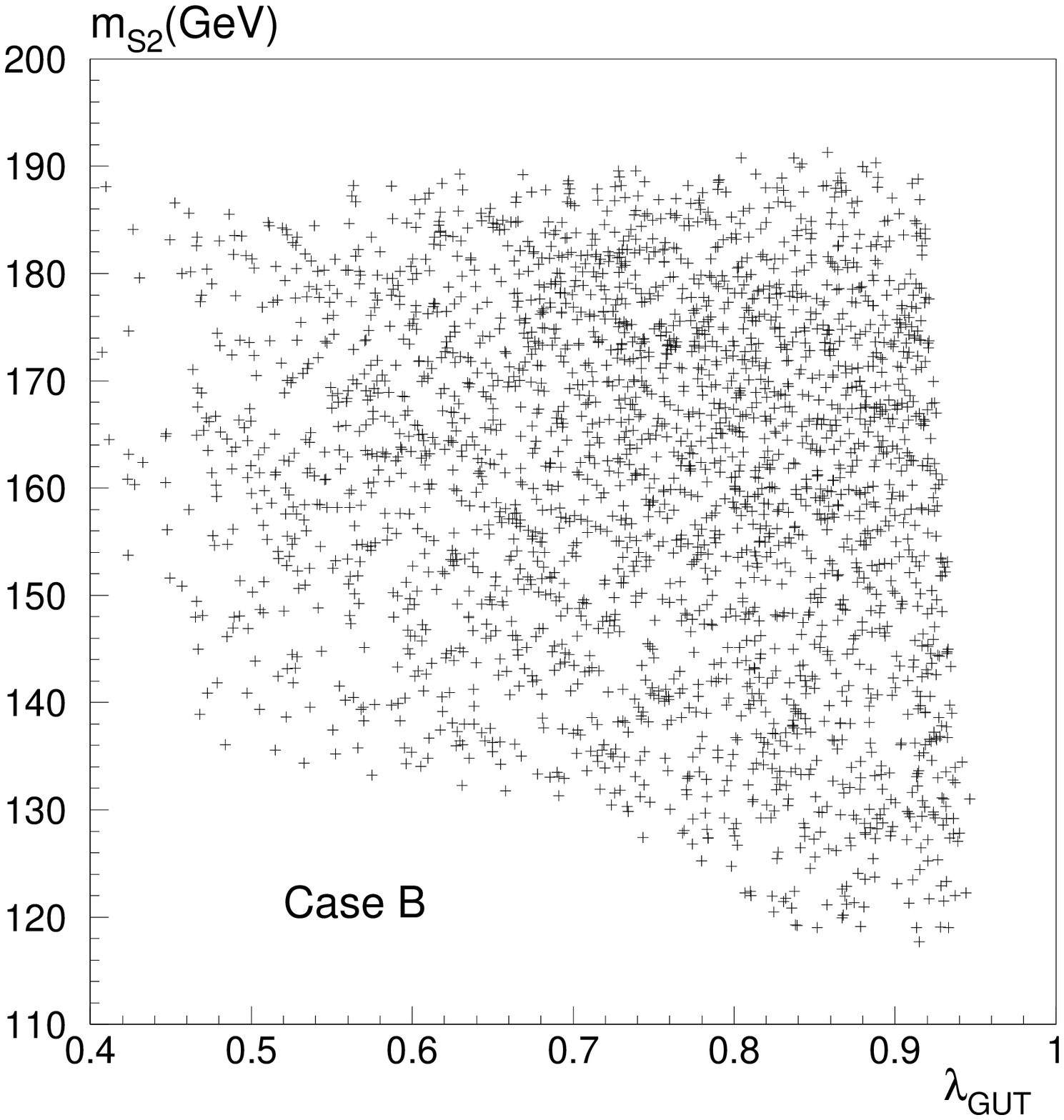}}
%\vspace*{1.cm}
\caption{The plot of the RG-improved mass at the one-loop level of $S_2$
against $\lambda_{\rm GUT}$, for $0 \leq \lambda_{\rm GUT} \leq 1.2$,
$m^2_{1_{\rm GUT}} = 0$, $0 < |m^2_{2_{\rm GUT}}|~\mbox{(TeV$^2$)} \leq 1$ and
$0 < |m^2_{3_{\rm GUT}}|~\mbox{(TeV$^2$)} \leq 1$ at the GUT scale.}
\label{b-ms2}
\end{center}
\end{figure}

\noindent
Fig. \ref{b-ms2}: \ The plot of the RG-improved mass at the one-loop level of $S_2$
against $\lambda_{\rm GUT}$, for $0 \leq \lambda_{\rm GUT} \leq 1.2$,
$m^2_{1_{\rm GUT}} = 0$, $0 < |m^2_{2_{\rm GUT}}|~\mbox{(TeV$^2$)} \leq 1$ and
$0 < |m^2_{3_{\rm GUT}}|~\mbox{(TeV$^2$)} \leq 1$ at the GUT scale.
\vskip 0.2 in

\begin{figure}[tbp]
\begin{center}
\vspace*{-0.1cm}
\mbox{\epsfxsize=8cm \epsfysize=8cm
\epsffile{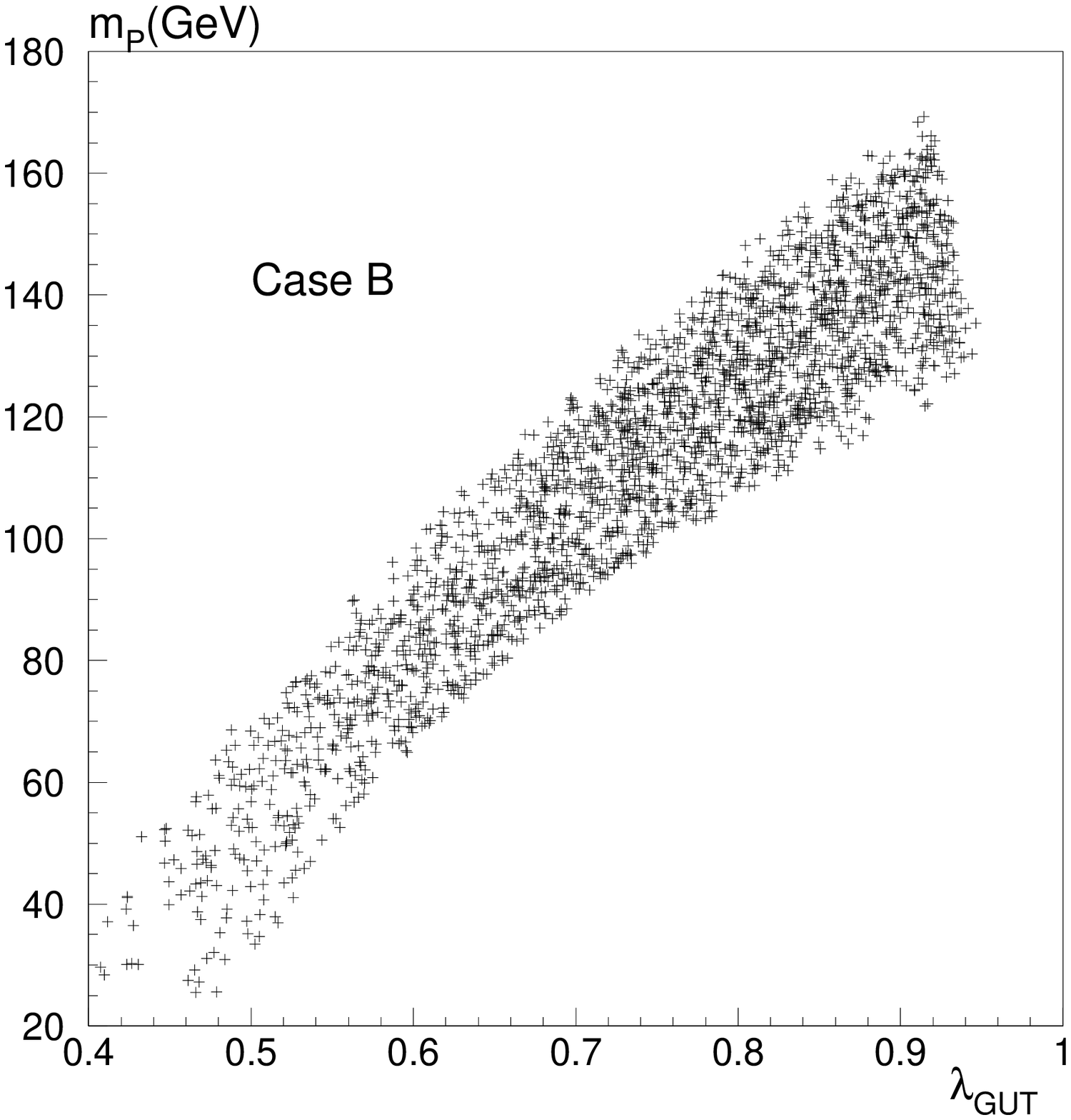}}
\caption{The plot of the RG-improved mass at the one-loop level of $P$
against $\lambda_{\rm GUT}$, for $0 \leq \lambda_{\rm GUT} \leq 1.2$,
$m^2_{1_{\rm GUT}} = 0$, $0 < |m^2_{2_{\rm GUT}}|~\mbox{(TeV$^2$)} \leq 1$ and
$0 < |m^2_{3_{\rm GUT}}|~\mbox{(TeV$^2$)} \leq 1$ at the GUT scale.}
\label{b-mp}
%\vspace*{-0.cm}
\end{center}
\end{figure}

\noindent
Fig. \ref{b-mp}: \ The plot of the RG-improved mass at the one-loop level of $P$
against $\lambda_{\rm GUT}$, for $0 \leq \lambda_{\rm GUT} \leq 1.2$,
$m^2_{1_{\rm GUT}} = 0$, $0 < |m^2_{2_{\rm GUT}}|~\mbox{(TeV$^2$)} \leq 1$ and
$0 < |m^2_{3_{\rm GUT}}|~\mbox{(TeV$^2$)} \leq 1$ at the GUT scale.
\vskip 0.2 in

\begin{figure}[h]
\begin{center}
%\vspace*{-0.1cm}
\mbox{\epsfxsize=8cm \epsfysize=8cm
\epsffile{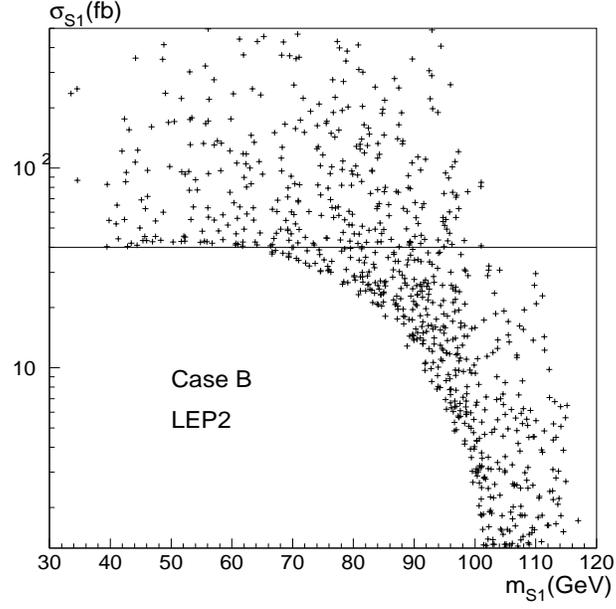}} \\
\vspace*{.5cm}
\mbox{\epsfxsize=8cm \epsfysize=8cm
\epsffile{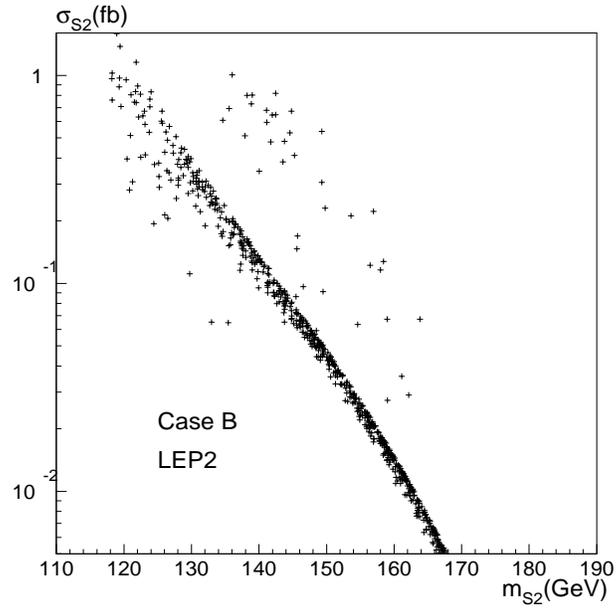}}
%\centerline{\includegraphics*[scale=0.5]{/@thesis/ch3fig/fig2a.eps}}
\caption{The plot against $m_{S_1}$ and $m_{S_2}$ of $\sigma_{S_1}$
and $\sigma_{S_2}$, respectively, at the one-loop level with the RG-improved
effective potential $V_{\rm eff}$ at LEP2 for $\sqrt{s} = 205.9 ~{\rm GeV}$
in the Case B.}
\label{b-lep206-ms1ms2}
%\vspace*{-10.3cm}
\end{center}
\end{figure}

\noindent
Fig. \ref{b-lep206-ms1ms2}: \ The plot against $m_{S_1}$ and $m_{S_2}$ of $\sigma_{S_1}$
and $\sigma_{S_2}$, respectively, at the one-loop level with the RG-improved
effective potential $V_{\rm eff}$ at LEP2 for $\sqrt{s} = 205.9 ~{\rm GeV}$
in the Case B.
\vskip 0.2 in

\begin{figure}[h]
\begin{center}
%\vspace*{-0.1cm}
\mbox{\epsfxsize=8cm \epsfysize=8cm
\epsffile{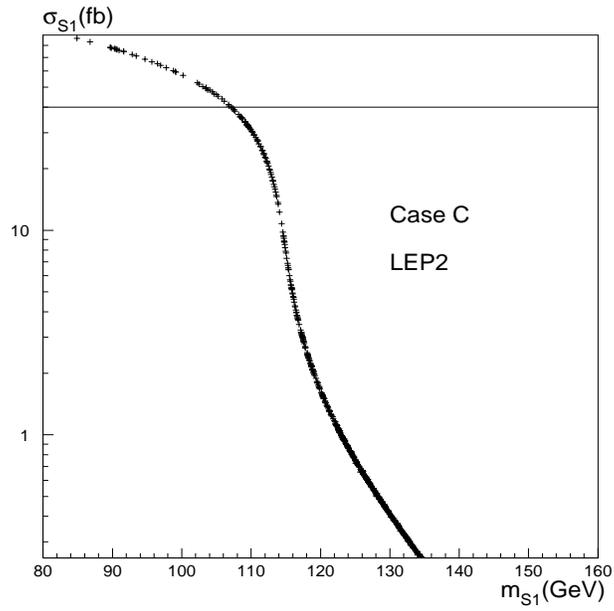}}
%\centerline{\includegraphics*[scale=0.5]{/@thesis/ch3fig/fig2a.eps}}
\caption{The plot of the largest $\sigma_{S_1}$
for given $m_{S_1}$ in the entire region of the parameter space
at the one-loop level with the RG-improved effective potential
$V_{\rm eff}$ at LEP2 for $\sqrt{s} = 205.9 ~{\rm GeV}$ in the Case C.}
\label{c-lep206-ms1}
%\vspace*{-10.3cm}
\end{center}
\end{figure}

\noindent
Fig. \ref{c-lep206-ms1}: \ The plot of the largest $\sigma_{S_1}$
for given $m_{S_1}$ in the entire region of the parameter space
at the one-loop level with the RG-improved effective potential
$V_{\rm eff}$ at LEP2 for $\sqrt{s} = 205.9 ~{\rm GeV}$ in the Case C.
%***********************************************************************************
%***********************************************************************************
% Figures
%***********************************************************************************
%***********************************************************************************

%\newpage
%Fig.1
%***********************************************************************************
%Fig.2
%***********************************************************************************
\newpage
%Fig.3
%***********************************************************************************
%Fig.4
%***********************************************************************************
%Fig.5
%***********************************************************************************
%Fig.6
%***********************************************************************************
%Fig.7
%\newpage
%***********************************************************************************
%Fig.8
%\newpage

%***********************************************************************************
\end{document}